\documentclass[conference]{IEEEtran}

\usepackage{cite}
\usepackage{amsmath,amssymb,amsfonts}
\usepackage{algorithmic}
\usepackage{graphicx}
\usepackage{textcomp}
\usepackage{xcolor}
\usepackage{balance}
\usepackage{eso-pic}
\def\BibTeX{{\rm B\kern-.05em{\sc i\kern-.025em b}\kern-.08em
    T\kern-.1667em\lower.7ex\hbox{E}\kern-.125emX}}

\begin{document}
\AddToShipoutPictureBG*{%
  \AtPageUpperLeft{%
    \put(\LenToUnit{\dimexpr(\paperwidth-\textwidth)/2\relax},\LenToUnit{-0.2in}){%
      \parbox[t]{\textwidth}{\scriptsize\centering
        Accepted to the IEEE ICDM International Workshop on AI for Personalization (WAIP 2026). To appear in the 2026 IEEE International Conference on Data Mining Workshops (ICDMW) proceedings.\\
        \copyright\,2026 IEEE. Personal use of this material is permitted. Permission from IEEE must be obtained for all other uses, in any current or future media, including reprinting/republishing this material for advertising or promotional purposes, creating new collective works, for resale or redistribution to servers or lists, or reuse of any copyrighted component of this work in other works.}}}}

\title{Dynamic Click Destinations: Personalizing Post\\Ad-Click Landing Experiences}

\author{\IEEEauthorblockN{Apurva Pathak, Yuya Shiraki, Iman Jenabzadeh, Nikhil Bhale,\\
Yuwen Tan, Shashank Rangarajan, Amirmehdi Sharifzad}
\IEEEauthorblockA{\textit{Meta Platforms}\\
Menlo Park, California, USA\\
\{apurvapathak, yshiraki, ijenab, nbhale, yuwent, sharanga, mehty\}@meta.com}
}

\maketitle

\begin{abstract}
Apps that monetize outbound (offsite) ad clicks have multiple post ad-click levers that influence user and business outcomes, including the landing page shown on the advertiser's site, whether the click resolves to a webpage or an app on the user's device, and whether the click resolves to a native in-product surface or an external website. We refer to the broader class of personalization problems over these levers as \emph{post ad-click destination automation}. Importantly, the optimal destination is heterogeneous across users, advertisers, placements, devices, and inferred intent, making static, one-size-fits-all destination policies suboptimal.

We present \emph{Dynamic Click Destinations} (DCD), a production personalization framework for post ad-click destination automation that selects a destination for each eligible offsite ad click to maximize \emph{Advertiser Value} (AV), a value-weighted conversion outcome. The DCD formulation is general and applies to any post-click destination decision where: (i) the system can choose among a set of destination options at click time, and (ii) downstream conversion feedback is available for learning and evaluation. We describe an evolution of learning objectives from (i) post-conversion destination inference under a single dominant destination, to (ii) two-task destination-conditioned conversion prediction, and finally to (iii) a single-task policy model that directly predicts which of two destinations is better for a given click. The final model is trained with an Advantage Actor Critic (A2C)-inspired soft-label construction derived from the logit difference of a higher-capacity destination-conditioned ``critic'' model, improving alignment with the serving-time decision boundary.

We evaluate DCD with a case study on \emph{web-vs-app destination automation}, where the per-click choice is between opening the advertiser's website (Web) and opening the advertiser's app on the user's device, when installed (App). Online A/B experiments at scale on a major online advertising platform show that per-click personalization with DCD outperforms both static routing policies---always sending eligible clicks to one or the other destination---with the single-task objective delivering the largest gains over earlier DCD generations. These results highlight post ad-click destination as an important and underexplored lever in conversion optimization and demonstrate that personalized destination automation can outperform one-size-fits-all destination policies.
\end{abstract}

\begin{IEEEkeywords}
post ad-click destination automation, landing destination, landing page personalization, web destination, app destination, personalization, ads optimization, contextual bandits, policy learning
\end{IEEEkeywords}

\section{Introduction}

Recommender and ranking systems typically optimize \emph{what} content or ad to show. In many consumer apps, however, a meaningful portion of the user and business outcome is determined \emph{after} a user clicks---particularly for offsite clicks that transition into a web context. We refer to the class of personalization problems over post-click destinations as \emph{post ad-click destination automation}, which spans several distinct levers, including:

\begin{itemize}
  \item \textbf{Landing page personalization}: choosing which page on the advertiser's site to load after a click (e.g., a generic homepage vs.\ a product detail page vs.\ a category page).
  \item \textbf{Web-vs-app destination personalization}: choosing whether to render the landing experience as a webpage in the post-click flow, or to route the click to an app on the user's device that preserves persistent state.
  \item \textbf{Native vs.\ external destination personalization}: choosing between an in-product surface (e.g., an in-app shop or business profile) and an external website for the same click intent.
\end{itemize}

This decision class arises across product categories, including \textbf{social media}, \textbf{messaging}, \textbf{news}, \textbf{e-commerce and shopping}, and emerging \textbf{AI-driven apps}, whenever an app mediates outbound clicks and can choose among multiple destination options. The optimal destination is rarely universal: each option has distinct trade-offs across user experience, conversion friction, and business value, and the better choice depends heavily on context such as user intent, advertiser type, and surface. For example, a furniture ad may benefit from a destination that supports a longer, research-heavy purchase journey, while a hair salon lead ad may benefit from a destination that lets users complete a short form and seamlessly return to the app. This kind of context-dependent treatment heterogeneity is the central object of study in the literature on \emph{uplift modeling} and \emph{heterogeneous treatment effect estimation} \cite{athey2016recursive,wager2018estimation,kunzel2019metalearners}, where the quantity of interest is the \emph{differential} effect of one option over another conditional on features rather than the absolute outcome under any single option.

These properties---per-click selection among actions with context-dependent outcomes, learning from logged feedback, and a value-based business objective---place post ad-click destination automation squarely in the setting of \emph{contextual bandits} and \emph{policy learning} \cite{li2010contextual,swaminathan2015counterfactual,agarwal2016making,dudik2011doublyrobust}, which learn decision rules from logged interaction data rather than predicting outcomes in isolation. In this work, we present \textbf{Dynamic Click Destinations (DCD)}, a framework that adopts this contextual decision-making perspective for post ad-click destination automation. DCD applies to any post-click destination decision where the system can select among multiple destinations at click time and observe downstream conversion feedback. We evaluate and deploy DCD at scale on a major online advertising platform with a case study on \emph{web-vs-app destination automation}---selecting at click time between a Web destination and an App destination, defined formally in Section~\ref{sec:web_vs_app}.

A natural modeling approach for this decision is to estimate destination-conditioned conversion probabilities $P(\mathrm{conv} \mid \mathrm{click}, a)$ separately for each destination $a$ and pick the higher-value option at serving time. Two-task formulations of this form, however, can be misaligned with the deployment decision: the policy depends on the \emph{relative advantage} of one destination over another, while the loss rewards calibration of absolute conversion propensity. Our central modeling contribution is a single-task formulation that targets this relative advantage directly. The construction is inspired by \emph{actor--critic} methods in reinforcement learning \cite{mnih2016a3c,sutton2018reinforcement,gauci2018horizon}, where policies are trained using advantage-style signals derived from value estimates: although DCD operates in a contextual bandit regime (no sequential state transitions), we derive dense supervision for the policy from the difference between destination-conditioned conversion logits produced by a larger ``critic'' model, and train a lightweight ``actor'' model aligned to the serving-time decision boundary.

We make three contributions:

\begin{enumerate}
  \item \textbf{A recommender framing of post ad-click destination automation.} We formalize landing destination as a per-click recommendation action over a set of available destination options, optimized for advertiser value. The framing is general and applies across landing-page personalization, web-vs-app destination personalization, and native-vs-external destination decisions.

  \item \textbf{An ML formulation aligned to the decision objective.} We describe an evolution from two-task destination-conditioned conversion prediction to a single-task objective that directly predicts which destination is better, using an \emph{Advantage Actor Critic (A2C)} inspired advantage-style label construction. The methods apply to any binary post-click destination decision and admit straightforward extensions to multi-option settings.

  \item \textbf{Online evidence at scale on web-vs-app destination automation.} We summarize a holdout demonstrating that flipping the default destination is globally harmful yet beneficial for high-intent segments, motivating personalization; and we report online gains from the A2C-inspired formulation over earlier generations for the web-vs-app destination decision.
\end{enumerate}

\section{Problem Setting}

For each eligible offsite ad click with context $x$ (user, ad, advertiser, placement/surface, device, time, etc.), the system chooses an action from a set of available post-click destinations:

\begin{equation}
a \in \mathcal{A}(x).
\end{equation}

Concrete instantiations of $\mathcal{A}(x)$ include:
\begin{itemize}
  \item Landing page selection: candidate landing pages on the advertiser's site (e.g., homepage, product detail page, or category page).
  \item Web-vs-app destination selection: $\mathcal{A}(x) = \{\mathrm{Web}, \mathrm{App}\}$, where Web denotes rendering the landing experience as a webpage and App denotes routing to an app on the user's device (Section~\ref{sec:web_vs_app}).
  \item Native vs.\ external destination: a native in-product surface (e.g., in-app shop or business profile) and an external website.
\end{itemize}

The user then experiences the landing environment and may convert (purchase, lead, sign-up, etc.) within an attribution window. The goal is to select the destination that maximizes expected advertiser value for each click.

For the remainder of this paper, we instantiate $\mathcal{A}(x) = \{\mathrm{Web}, \mathrm{App}\}$, defined in Section~\ref{sec:web_vs_app}, since this case study grounds our experimental evaluation.

\subsection{Web and app destinations}
\label{sec:web_vs_app}

In our case study, the per-click choice is between two post-click destinations:

\begin{itemize}
  \item \emph{Web}: the landing page is rendered as a webpage within the post-click flow, using a native web-viewer. This is the default destination prior to DCD.
  \item \emph{App}: the click is routed to an app on the user's device that already holds the user's persistent state, including authentication, long-lived cookies, autofill, and stored payment credentials. This is the alternative destination that DCD may route to.
\end{itemize}

The two destinations have distinct trade-offs. Web provides a consistent, predictable post-click flow with lower context-switching cost and tight integration with the host app. App preserves continuity with the user's existing on-device state, which can reduce friction for tasks like authentication, checkout, or autofill on advertiser sites in certain use cases. These selective benefits motivate routing some click contexts to App when the expected conversion advantage outweighs the additional context switch, rather than adopting App as a blanket default.

\subsection{Objective and evaluation metric: Advertiser Value}
\label{sec:ads_value}

We optimize and evaluate DCD using \emph{Advertiser Value (AV)}, a value-based outcome that captures both the volume of conversion and the business value of those conversions to advertisers. We define AV as a conversion value weighted sum over conversion events:

\begin{equation}
\mathrm{AV} \;=\;
\sum_{i \in \text{clicks}} v_{i}\cdot
\mathbf{1}\{\text{conversion} \text{ occurs for click } i\},
\label{eqn:ads_value}
\end{equation}

where $v_{i}$ is the advertiser value associated with conversion for click $i$ (e.g., reported value, modeled value, or a proxy). In online experiments, we report relative lift in AV (normalized to control).

\subsection{Why static routing fails: heterogeneous treatment effects}

A key property of post ad-click destination automation is heterogeneity: the better destination depends on click context. In a large-scale holdout that flipped the default destination from Web to App for eligible offsite ad clicks, global \emph{Advertiser Value (AV)} regressed by \textbf{1.05\%} relative to the always-Web baseline. However, we observed a \textbf{0.1\% AV gain} on the segment of ads optimizing for down-funnel objectives such as \emph{Purchase}. Detailed experimentation results are provided in Section~\ref{sec:experiments}. This pattern suggests that neither destination dominates globally, and motivates a personalized per-click decision rule:

\begin{quote}
Route to the alternative destination only when it is expected to outperform the default for the specific click context; otherwise, keep the default.
\end{quote}

\section{Method}
\label{sec:method}

This section describes the evolution of model formulations culminating in a single-task model aligned to the decision: ``which post-click destination is better for this click?'' We present the formulation for the binary destination case study (Web vs.\ App), since it grounds our experimental evaluation; the formulation generalizes to other binary post-click destination decisions (e.g., native vs.\ external surface) and admits straightforward extensions to multi-option settings (e.g., landing page personalization with $k$ candidate pages). Across all generations, we use the \emph{Deep Learning Recommendation Model (DLRM)} \cite{naumov2019deep} architecture to model click context features. The detailed model architecture is outside the scope of this paper.

A key enabler for learning and evaluation is that the system logs (i) the \emph{landing destination} chosen for every eligible ad click and (ii) the \emph{conversion destination} for every observed conversion event.

\subsection{Generation 1: Post-conversion destination inference without alternative-destination click prior}
\label{sec:gen1}

Before DCD, Web was the default landing destination, leaving limited data to directly estimate outcomes under the alternative destination. A first-generation approach used post-conversion data observed under the default destination and labeled examples based on the conversion's destination, training only on clicks that landed in the default.

We then used the model to infer whether a click should be routed to the alternative destination based on the probability that the eventual conversion (if any) would occur in the alternative, given that the click initially landed in the default:

\begin{multline}
\text{If } P(\texttt{conv\_dest} = \texttt{App} \mid \texttt{click}, \texttt{land\_dest}=\texttt{Web}) \\
> \tau \text{ then } \texttt{land\_dest}=\texttt{App} \\
\text{ else } \texttt{land\_dest}=\texttt{Web},
\label{eq:gen1_rule}
\end{multline}

where $land\_dest$ and $conv\_dest$ refer to the landing destination and conversion destination respectively and the threshold $\tau$ regulates the volume of clicks routed to the alternative destination.

While this bootstrapped initial routing, it did not directly model the counterfactual question of which landing destination would yield higher conversion probability/value for a given click. This particular bootstrapping strategy relies on a property specific to the web-vs-app case: even when a click lands in Web, the conversion may occur in App (e.g., if the user navigates from one to the other before converting), so the post-conversion destination provides a usable proxy label. For other post-click destination automation use cases where the alternative destination is not reachable from the default (and thus the post-conversion destination is uninformative), an analogous bootstrapping strategy is to randomly route a small fraction of traffic to each candidate destination, generating the exploration data needed to learn the relative advantage in subsequent generations.

\subsection{Generation 2: Two-task destination-conditioned conversion prediction}
\label{sec:gen2}

After deploying destination routing, the system observes click traffic under both destinations. This enables learning destination-conditioned \emph{conversion rates} (CVR), the per-click probability of conversion under each candidate destination:

\begin{multline}
\texttt{app\_cvr} = \\
P(\texttt{conversion} \mid \texttt{click}, \texttt{land\_dest}=\texttt{App}),
\label{eq:papp}
\end{multline}

\begin{multline}
\texttt{web\_cvr} = \\
P(\texttt{conversion} \mid \texttt{click}, \texttt{land\_dest}=\texttt{Web}).
\label{eq:pweb}
\end{multline}

A decision rule chooses the alternative destination when its predicted conversion probability sufficiently exceeds that under the default:

\begin{multline}
\text{If } \texttt{app\_cvr}
>
\tau \cdot \texttt{web\_cvr} \\
\text{ then } \texttt{land\_dest}=\texttt{App}
\text{ else } \texttt{land\_dest}=\texttt{Web},
\label{eq:gen2_rule}
\end{multline}

where the threshold $\tau$ regulates the volume of clicks routed to the alternative destination.

This formulation is closer to the ideal, but introduces a learning misalignment: two-task training rewards features that predict conversion likelihood overall, not necessarily features that best discriminate between destinations.

\subsection{Generation 3: Directly predict ``alternative destination is better''}
\label{sec:gen3}

To align modeling with the decision objective, we shift to a single prediction target. For the web-vs-app case study, this is:

\begin{equation}
P(\texttt{App better than Web} \mid \texttt{click}).
\label{eq:single_task_target}
\end{equation}

In the general DCD formulation with two destinations $a_0$ (default) and $a_1$ (alternative), the corresponding target is
\begin{equation*}
P(a_1 \text{ better than } a_0 \mid \text{click}).
\end{equation*}
This directly optimizes the decision boundary and encourages feature learning that captures relative destination advantage rather than absolute CVR.

\subsubsection{Advantage Actor Critic (A2C) inspired label construction using an upstream critic model}
\label{sec:a2c_labels}

We use a larger ``critic'' model that outputs logits for destination-conditioned conversion probabilities:

\begin{itemize}
    \item $\ell_{\texttt{App}}(\texttt{click})$: logit for $P(\texttt{conversion} \mid \texttt{click}, \texttt{land\_dest}=\texttt{App})$
    \item $\ell_{\texttt{Web}}(\texttt{click})$: logit for $P(\texttt{conversion} \mid \texttt{click}, \texttt{land\_dest}=\texttt{Web})$
\end{itemize}

We then define soft labels for the single-task model via a sigmoid of the logit difference:

\begin{equation}
 y(\texttt{click}) = \sigma\!\left(\alpha \cdot \bigl(\ell_{\texttt{App}}(\texttt{click}) - \ell_{\texttt{Web}}(\texttt{click})\bigr)\right),
 \label{eq:soft_label}
\end{equation}

where $\sigma(\cdot)$ is the sigmoid function and $\alpha$ is a positive scaling hyperparameter applied to the logit difference, since the raw difference is typically very small in magnitude. Intuitively, if the critic predicts the alternative destination is much better than the default, $y \rightarrow 1$; in the opposite case, $y \rightarrow 0$. This is A2C-inspired in the sense that the supervised target is derived from an advantage-like quantity (relative logit difference) rather than training two separate CVR heads and subtracting them at inference.

\subsubsection{Decision rule}
\label{sec:decision_rule}

At serving time, the DCD model predicts $P(\texttt{App better} \mid \texttt{click})$ and the policy chooses the alternative destination when this probability exceeds the threshold:

\begin{multline}
    \text{If } P(\texttt{App better} \mid \texttt{click}) > \tau \\
    \text{then } \texttt{land\_dest} = \texttt{App} \text{ else } \texttt{land\_dest} = \texttt{Web}
    \label{eq:gen3_rule}
\end{multline}

where the threshold $\tau$ regulates the volume of clicks routed to the alternative destination.

\section{Experiments}
\label{sec:experiments}

We evaluate the Dynamic Click Destinations (DCD) framework with a case study on web-vs-app destination automation, using \emph{online A/B tests} at scale on a major online advertising platform, across surfaces where eligible offsite ad clicks can be routed at click time to either of the two destinations defined in Section~\ref{sec:web_vs_app}. Each experiment runs for three weeks, with the test and control cells each sized at 10\% of overall platform user traffic. The platform serves hundreds of billions of ad impressions and billions of ad clicks per day in aggregate; the offsite subset eligible for DCD routing accounts for tens of billions of impressions and over a billion clicks per day. Each three-week experiment cell therefore observes more than 100 billion eligible impressions and several billion eligible offsite clicks, providing ample statistical power for the AV lifts reported below. We report results using \emph{Advertiser Value (AV)} (Section~\ref{sec:ads_value}), as relative lift normalized to an ``always-Web'' (default) control (0.00\%).

\begin{table}[!t]
\renewcommand{\arraystretch}{1.3}
\caption{Advertiser Value (AV) lift vs.\ always-Web (default) baseline for the web-vs-app destination automation case study.}
\label{tab:ads_value_results}
\centering
\begin{tabular}{p{5cm} r}
\hline
\textbf{Policy / Model} & \textbf{Advertiser Value} \\
\hline
\textbf{Always-Web (baseline)}: all eligible clicks land in the default destination & 0.00\% \\
\textbf{Always-App}: all eligible clicks land in the alternative destination & $-1.05\%$ \\
\textbf{Generation 1 DCD}: Post-conversion destination inference without alternative-destination click prior & $+0.06\%$ \\
\textbf{Generation 2 DCD}: Two-task destination-conditioned CVR prediction & $+0.42\%$ \\
\textbf{Generation 3 DCD}: Single-task ``alternative destination is better'' (A2C-inspired soft-label) & $+0.54\%$ \\
\hline
\end{tabular}
\end{table}

\subsection{Online performance}
\label{sec:online_perf}

Table~\ref{tab:ads_value_results} summarizes the AV impact of static routing and successive DCD generations on the web-vs-app case study. Two takeaways stand out.

First, \emph{flipping the default destination is globally harmful but selectively beneficial}: routing all eligible clicks to the alternative (always-App) yields a \(-1.05\%\) AV regression vs.\ the original default (always-Web). However, on the segment of ads optimizing for down-funnel objectives such as \emph{Purchase}, the same always-App policy yields a \(+0.1\%\) AV gain. Together, these results confirm that a one-size-fits-all policy is suboptimal and motivate per-click personalization.

Second, \emph{personalization improves AV and improves with better objectives}. The three DCD generations show monotonic gains relative to the always-Web baseline: Generation~1 (post-conversion destination inference) yields \(+0.06\%\) AV, Generation~2 (two-task destination-conditioned CVR prediction) improves to \(+0.42\%\) AV, and Generation~3 (single-task ``alternative-destination-is-better'' with Advantage Actor Critic (A2C)-inspired soft labels) performs best at \(+0.54\%\) AV, outperforming the two-task approach by \(+0.12\%\) AV absolute.

\subsection{Iteration velocity: why the A2C formulation helped beyond AV}
\label{sec:iteration_velocity}

The benefit of the A2C-inspired single-task model was not limited to higher online AV; it also reduced the execution cost of iteration by improving the reliability of offline evaluation.

With the two-task model, standard offline metrics such as Normalized Entropy (NE) \cite{he2014practical} were often not well correlated with online business outcomes such as AV. A key reason is objective mismatch: each head is trained to predict CVR under a fixed destination, rather than to directly optimize the decision boundary of choosing the better destination. As a result, offline improvements could reflect better calibration or CVR prediction accuracy while failing to translate into better destination selection decisions, increasing the dependence on online tests to validate model changes.

With the A2C-inspired single-task model, the prediction target is explicitly aligned with the decision (``alternative destination is better than default''), and we observed that offline metrics became directionally correlated with online AV. Concretely, we tracked the sign of each candidate iteration's offline NE delta against the sign of its online AV delta across the candidate iterations evaluated during DCD development; under the A2C-inspired formulation the two were consistently aligned, whereas under the two-task formulation the signs frequently disagreed. This alignment made offline experimentation more useful for prioritizing candidate iterations, reducing iteration time and increasing confidence that offline wins would carry through online.

\subsection{DCD as an ads-ranking feature (eCVR model)}
\label{sec:dcd_as_feature}

Beyond direct policy improvements from destination routing optimization, we also evaluated the DCD output---the predicted post-click destination for an ad click---as an input feature to the ads ranking model that predicts \(\mathrm{eCVR} = P(\texttt{conversion} \mid \texttt{impression})\). In this analysis, the DCD destination feature ranked among the top 3 features by feature importance. This suggests the personalized destination decision captures strong signal correlated with downstream conversion propensity, and that DCD can create value both through better post-click routing and through improved ranking when used as an input to downstream models.

\section{Conclusions}
\label{sec:conclusions}

We presented Dynamic Click Destinations (DCD), a personalization framework for post ad-click destination automation. Our case study on web-vs-app destination automation, evaluated at scale on a major online advertising platform, shows that flipping the default destination is globally harmful (\(-1.05\%\) AV) while per-click personalization with DCD yields monotonic improvements across generations, peaking at \(+0.54\%\) AV with the Advantage Actor Critic (A2C)-inspired single-task formulation (Section~\ref{sec:online_perf}). Aligning the training objective with the destination decision also improved iteration efficiency by making offline metrics directionally correlated with online AV (Section~\ref{sec:iteration_velocity}), and the predicted destination ranked among the top 3 features by importance when used as an input to the eCVR ranking model (Section~\ref{sec:dcd_as_feature}).

Although this paper grounded DCD in a web-vs-app destination case study, the framework applies broadly to other post-click destination decisions, including landing page personalization and native-vs-external destination selection, across product categories such as social media, messaging, news, and shopping apps.

\balance
\bibliographystyle{IEEEtran}
\bibliography{sample-base}

\end{document}